\documentclass[10pt]{article}

\usepackage{mathptm}                               
\usepackage{rotating}
\usepackage{epsfig}
\usepackage{amsmath}
\usepackage{amsfonts}
\usepackage{multirow} 
\usepackage{subfigure}
\usepackage{cite}

\begin{document}
\def\reals{\textbf{R}}


\title{Distributed NEGF Algorithms for the Simulation 
of Nanoelectronic Devices with Scattering}
\date{}

\author{
Stephen Cauley, Mathieu Luisier, Venkataramanan Balakrishnan, \\
Gerhard Klimeck, and Cheng-Kok Koh\\
 School of Electrical and Computer Engineering \\
 Purdue University, West Lafayette, IN 47907-2035 \\
}   

\maketitle
\begin{abstract}
Through the Non-Equilibrium Green's Function (NEGF) 
formalism, quantum-scale device simulation 
can be performed with the inclusion of 
electron-phonon scattering.
However, the simulation of realistically
sized devices under the NEGF formalism 
typically requires prohibitive amounts of 
memory and computation time.
Two of the most demanding computational problems 
for NEGF simulation involve mathematical operations with 
structured matrices called semiseparable matrices.
In this work, we present parallel approaches for these 
computational problems which allow for efficient distribution
of both memory and computation based upon the underlying
device structure.
This is critical when simulating realistically sized devices due to 
the aforementioned computational burdens.
First, we consider determining a distributed compact representation
for the retarded Green's function matrix $G^{R}$.
This compact representation is exact and
allows for any entry in the matrix to be generated 
through the inherent semiseparable structure.
The second parallel operation allows for the computation of 
electron density and current characteristics for the device.
Specifically, matrix products between the distributed representation
for the semiseparable matrix $G^{R}$ and the
self-energy scattering terms in $\Sigma^{<}$ 
produce the less-than Green's function $G^{<}$.
As an illustration of the computational efficiency of our approach, 
we stably generate the mobility for nanowires with 
cross-sectional sizes of up to $4.5$nm, 
assuming an atomistic model with scattering.
\end{abstract}

\section{Introduction}
\label{ss:intro}
In the absence of electron-phonon scattering, 
the problem of computing density of states and
transmission through the NEGF formalism reduces to a mathematical 
problem of finding select entries from the inverse of a typically
large and sparse block tridiagonal matrix.
Although there has been research into numerically stable
and computationally efficient serial computing 
algorithms~\cite{anant}, 
when analyzing certain device geometries this 
type of method will result in prohibitive amounts of computation
and memory consumption. 
In~\cite{cauley:07} a parallel divide-and-conquer algorithm (PDIV)
 was shown to be effective for NEGF based simulations.
Two applications were presented, the atomistic level simulation 
of silicon nanowires and the two-dimensional simulation of nanotransistors.
Alternative, serial computing, NEGF based approaches such as~\cite{darve:08} 
rely on specific problem structure (currently not capable
of addressing atomistic models), with computation limitations
that again restrict the size of simulation.
In addition, the ability to compute information 
needed to determine current characteristics for devices
has not been demonstrated with methods such as~\cite{darve:08}.
The prohibitive computational properties associated with
NEGF based simulation has prompted a transition to 
wave function based methods, such as those presented in~\cite{tuttle_01}.
Given an assumed basis structure
the problem translates from calculating select entries 
from the inverse of a matrix to solving large sparse systems
of linear equations.  This is an attractive alternative because 
solving large sparse systems
of equations is one of the most well studied problems in applied mathematics
and physics.  In addition, popular algorithms such as 
UMFPACK~\cite{UMFPACK} and SuperLU~\cite{superlu99} have been constructed to algebraically 
(based solely on the matrix) exploit problem specific structure in an 
attempt to minimize the amount of computation.
The performance of these algorithms for the wave function based analysis
of several silicon nanowires has been examined in~\cite{boykin_08}.
However, for many devices of interest 
wave function methods are currently unable to address
more sophisticated analyses that involve electron-phonon scattering.
Thus, there remains a strong need to further develop NEGF based
algorithms for the simulation of realistically sized devices
 considering these general modeling techniques.

When incorporating the effects of scattering into NEGF
based simulation, it becomes necessary
to determine the entire inverse of the coefficient
matrix associated with the device. 
  This substantially increases both the computational and 
memory requirements.
There are a number of theoretical results describing the
structure of the inverses of block tridiagonal and block-banded
matrices.  Representations for the inverses of tridiagonal, banded,
and block tridiagonal matrices can be found in~\cite{bowden, bukhberger:mathphysics73, 
godfrin, meurant,oohashi:tru78, torii:osaka66}.  It
has been shown that the inverse of a tridiagonal matrix can be
compactly represented by two sequences $\{u_i\}$ and
$\{v_i\}$~\cite{asplund:E,asplund:S,romani,nabben}.  This result was
extended to the cases of block tridiagonal and banded matrices
in~\cite{romani:add,rozsa:add,rozsa:add1}, where the $\{u_i\}$ and
$\{v_i\}$ sequences generalized to matrices $\{U_i\}$ and $\{V_i\}$.
Matrices that can be represented in this fashion are
more generally known as semiseparable matrices~\cite{rozsa86,rozsa:add1}.
Typically, the computation of parameters $\{u_i\}$ and $\{v_i\}$ suffers
from numerical instability even for modest-sized problems~\cite{concus}.  
It is well understood that for matrices arising in many physical
applications the $\{u_i\}$ and $\{v_i\}$ sequences grow
exponentially~\cite{demko84:decay,nabben} with the index $i$.
One approach that has been successful in
ameliorating these problems, for the tridiagonal case, is the generator 
approach shown in~\cite{block_precond}.  Here, ratios for sequential 
elements of the $\{u_i\}$ and $\{v_i\}$ sequences
are used as the generators for the inverse of a 
tridiagonal matrix.  Such an approach is numerically stable for matrices
 of very large sizes.
 The extension of this generator approach to the general
block-tridiagonal matrices was discussed by the same authors
in~\cite{meurant}.  The authors used the block factorization of the
original block-tridiagonal matrix to construct a block Cholesky
decomposition of its inverse.  

A generator based approach for inversion,
 typically referred to as the
Recursive Green's Function (RGF) algorithm, was introduced 
in~\cite{anant} for NEGF based simulation.
It is important to note that the method of~\cite{anant} 
requires $3\times$ less memory to compute only the density of states
and transmission (in the absence of scattering),
when compared to the complete generator representation.
In this work, we extend the approach of~\cite{cauley:07}
to consider the computation of a distributed 
generator representation for the inverse 
of the block tridiagonal coefficient matrix.  
We then demonstrate how the distributed generator representation
allows for the efficient computation of the electron density and current 
characteristics of the device.
Our parallel algorithms facilitate
the simulation of realistically sized devices
 by utilizing additional 
computing resources to efficiently divide both the computation
time and memory requirements.
As an illustration, 
we stably generate the mobility for 
$4.5$nm cross-section nanowires 
assuming an atomistic model with scattering.

\section{Inverses of Block Tridiagonal Matrices}
\label{ss:blkt_inv}
A block-symmetric matrix $K$ is block tridiagonal if it has the form 
\begin{align}
K = \begin{pmatrix}
 A_1&-B_1\\
 -B_1^{T}&A_2&-B_2\\
 &\ddots&\ddots&\ddots\\
 &&-B_{N_{y}-2}^{T}&A_{N_{y}-1}&-B_{N_{y}-1}\\
 &&&-B_{N_{y}-1}^{T}&A_{N_{y}}\\
 \end{pmatrix},
 \end{align}
where each $A_{i},B_{i} \in \mathbb{R}^{N_{x}\times N_{x}}$.  Thus $K 
\in \mathbb{R}^{N_{y}N_{x}\times N_{y}N_{x}}$, with $N_{y}$ diagonal 
blocks of size $N_{x}$ each.
We will use the notation $K = \mathrm{tri}(A_{1:N_{y}}, B_{1:N_{y}-1})$
to represent such a block tridiagonal matrix.
The NEGF based simulation of nanowires using the
$sp3d5s*$ atomistic tight-binding model with electron-phonon
scattering has been demonstrated in~\cite{mLU_08}.
The block tridiagonal coefficient matrix for simulation is 
constructed in the following way:
\[
K = \left( EI - H - \Sigma^R_R - \Sigma^R_L - \Sigma^R_S \right).
\]
Here, $E$ is the energy of interest, $H$ is the Hamiltonian 
containing atomistic interactions, and $\Sigma^R_L$, 
 $\Sigma^R_R$, and $\Sigma^R_S$ are the left and right 
boundary conditions and self energy scattering terms 
respectively.
In order to calculate the current characteristics
for the device we must first form the retarded Greens 
Function using the fact that $K G^{R} = I$.
A standard numerically stable mathematical representation 
for the inverse of this block tridiagonal matrix 
is dependent on two sequences of generator matrices
$\{g^{\underleftarrow{R}}_{i}\}$,$\{g^{\overrightarrow{R}}_{i}\}$.
Here, the terms $\underleftarrow{R}$ and $\overrightarrow{R}$
correspond to the
forward and backward propagation through the device.
Specifically, we can use the diagonal blocks of the 
inverse $\{D_{i}\}$ and the generators 
to describe the inverse a block tridiagonal matrix
$K$ in the following manner:

\begin{align}
\label{eqn:K_ratio}
G^R =
\begin{pmatrix}
D_{1}&D_{1}g^{\overrightarrow{R}}_{1}&\cdots &D_{1}\prod\limits_{k = 1}^{N_{y}-1}g^{\overrightarrow{R}}_{k}\\
g^{\underleftarrow{R}}_{1}D_{1}&D_{2}&\cdots &D_{2}\prod\limits_{k = 2}^{N_{y}-1}g^{\overrightarrow{R}}_{k}\\
\vdots&\vdots&\ddots&\vdots\\
(\prod\limits_{k = N_{y}-1}^{1}g^{\underleftarrow{R}}_{k})D_{1}&(\prod\limits_{k = N_{y}-1}^{2}g^{\underleftarrow{R}}_{k})D_{2}&\cdots &D_{N_{y}}
\end{pmatrix}.
\end{align}

\noindent Where the diagonal blocks of the inverse, $D_{i}$, and the generator
 sequences satisfy the following 
relationships:
\begin{equation}
\label{eqn:uv}
\begin{array}{l}
g^{\underleftarrow{R}}_{1}=A_{1}^{-1}B_{1}, \\
g^{\underleftarrow{R}}_{i}=\left(A_{i}-B_{i-1}^{T}g^{\underleftarrow{R}}_{i-1}\right)^{-1}B_{i}, \quad
i=2,\ldots,N_{y}-1, \\
\\
g^{\overrightarrow{R}}_{N_{y}-1}=B_{N_{y}-1}A_{N_{y}}^{-1}, \\
g^{\overrightarrow{R}}_{i}=B_{i}\left(A_{i+1}-g^{\overrightarrow{R}}_{i+1}B_{i+1}^{T}\right)^{-1}, \quad  i=N_{y}-2,\ldots,1,
\\
\\
D_{1}= \left(A_{1}-g^{\overrightarrow{R}}_{1}B_{1}^{T}\right)^{-1}, \\
D_{i+1} = \left(A_{i+1} - g^{\overrightarrow{R}}_{i+1}B_{i+1}^{T}\right)^{-1} \left(I + B_{i}^{T}D_{i}g^{\overrightarrow{R}}_{i}\right),~~~~ i =1,...,N_{y}-2,
\\
D_{N_{y}} = A_{N_{y}}^{-1}\left(I + B_{N_{y}-1}^{T}D_{N_{y}-1}g^{\overrightarrow{R}}_{N_{y}-1}\right). 
\end{array}
\end{equation}
The time complexity associated with determining the parametrization of $G^R$ by
the above approach is $O(N_{x}^{3}N_{y})$, with a memory requirement of $O(N_{x}^{2}N_{y})$.

\subsection{Alternative Approach for Determining the Compact Representation}
\label{ss:toy_ex2}
It is important to note that if the block tridiagonal portion of 
$G^R$ is known, the generator sequences $g^{\overrightarrow{R}}$ and $g^{\underleftarrow{R}}$ 
can be extracted directly, i.e. without the use 
of entries from $K$ through the generator expressions~(\ref{eqn:uv}).
Examining closely the block tridiagonal portion of $G^R$
we find the following relations:
\begin{equation}
\begin{array}{l}
D_{i} g^{\overrightarrow{R}}_{i} = P_{i}~\implies g^{\overrightarrow{R}}_{i} = D_{i} ^ {-1} P_{i}, \quad
i=1,\ldots,N_{y}-1, \\
\\
g^{\underleftarrow{R}}_{i} D_{i} = Q_{i}~\implies g^{\underleftarrow{R}}_{i} = Q_{i} D_{i} ^ {-1}, \quad
i=1,\ldots,N_{y}-1, \\
\end{array}
\label{eqn:comp_ratio}
\end{equation}
where $P_{i}$ denotes the $(i, i+1)$ block entry of $G^R$
and $Q_{i}$ denotes the $(i+1, i)$ block entry of $G^R$.
Therefore, by being able to produce the block tridiagonal portion of $G^R$
we have all the information that is necessary to compute
the compact representation.

As was alluded to in Section~\ref{ss:intro}, direct techniques 
for simulation of realistic devices often
require prohibitive memory and computational requirements.
To address these issues we offer a parallel divide-and-conquer approach 
in order to construct the compact representation for $G^R$, 
i.e. the framework allows for the parallel inversion of the 
coefficient matrix.  Specifically, we introduce
an efficient method for computing the block tridiagonal 
portion of $G^R$ in order to exploit the process
demonstrated in~(\ref{eqn:comp_ratio}).

\section{Parallel Inversion of Block Tridiagonal Matrices}
\label{s:pinv_blktri}

\begin{figure*}
\begin{center}
\leavevmode
\epsfxsize=5in
\epsfbox{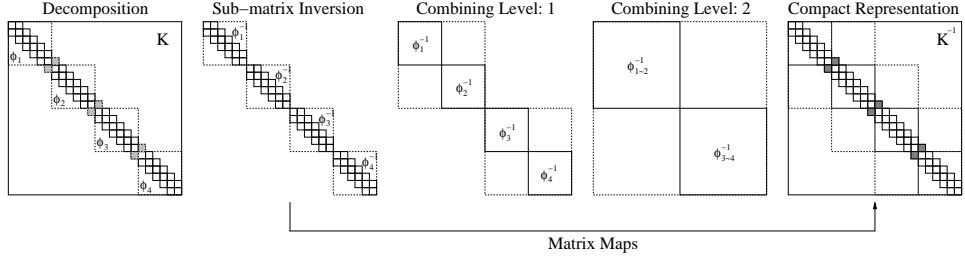}
\caption{Decomposition of block tridiagonal matrix $K$ into four sub-matrices, where the shaded blocks
correspond to the bridge matrices.  The two combining levels follow the individual sub-matrix inversions,
where $\phi_{i \sim j}^{-1}$  represents the inverse of divisions $\phi_{i}$ through $\phi_{j}$ from the matrix $K$.
Matrix mappings will be used to capture the combining effects and allow for the direct 
computation of the block tridiagonal portion of $G^R$.}
\label{fig:decomp_over}
\end{center}
\end{figure*}

The compact representation of $G^R$ can be computed 
in a distributed fashion by first creating several smaller
 sub-matrices $\phi_{i}$.
That is, the total number of blocks for the matrix $K$ 
are divided as evenly as possible amongst the sub-matrices.
After each individual sub-matrix inverse
has been computed they can be combined in a Radix-2 fashion 
using the matrix inversion lemma from linear algebra.
Figure~\ref{fig:decomp_over} shows both the decomposition and 
the two combining levels needed to form the block tridiagonal 
portion of $G^R$, assuming $K$ has been divided into four
sub-matrices.
In general, if $K$ is separated into $p$ sub-matrices
there will be $\log p$ combining levels with a total of $p - 1$
combining operations or ``steps''.
The notation $\phi_{i\sim j}^{-1}$ is introduced to represent  
the result of any combining step, through the use of the matrix inversion lemma.
For example, $\phi_{1\sim 2}^{-1}$ is the inverse of a matrix
comprised of the blocks assigned to both $\phi_1$ and $\phi_2$. 
It is important to note that using the
matrix inversion lemma repeatedly to join
sub-matrix inverses will result in a prohibitive
amount of memory and computation for large simulation problems.
This is due to the fact that at each combining step
all entries would be computed and stored.
Thus, the question remains on the
most efficient way to produce the block tridiagonal portion 
of $G^R$, given this general decomposition scheme for the matrix $K$.

In this work, we introduce a mapping scheme to transform compact
 representations of smaller matrix inverses
into the compact representation of $G^R$.
The algorithm is organized as follows: 

\begin{itemize}
\item Decompose the block tridiagonal matrix $K$ into $p$ smaller block 
tridiagonal matrices.
\item Assign each sub-matrix to an individual CPU.
\item Independently determine the compact representations associated 
with each sub-matrix.
\item Gather all information that is needed to map the sub-matrix compact 
representations into the compact representation for $G^R$.
\item Independently apply the mappings to produce a portion of the
compact representation for $G^R$ on each CPU.
\end{itemize}

\noindent The procedure described above results in a ``distributed compact
representation''  allowing for reduced memory and computational requirements.
Specifically, each CPU will eventually be responsible for the elements
from both the generator sequences and diagonal blocks that correspond to the initial
decomposition (e.g. if $\phi_1$ is responsible for blocks
$1, 2,~\mathrm{and}~3$ from the matrix $K$, the mappings will allow for the computation
of $g^{\overrightarrow{R}}_{1\ldots3},~g^{\underleftarrow{R}}_{1\ldots3},~\mathrm{and}~D_{1\ldots3}$).

In order to derive the mapping relationships needed to produce
a distributed compact representation, it is first necessary to 
analyze how sub-matrix inverses
can be combined to form the complete inverse.
Consider the decomposition of the block tridiagonal matrix $K$
into two block tridiagonal sub-matrices and a correction term, demonstrated below:

\begin{align*}
K &= \underbrace{\begin{pmatrix}
 \phi_1\\
 &\phi_2
 \end{pmatrix}}_{\tilde{K}}~~+~~XY,\\
\phi_{1} &= \mathrm{tri}(A_{1:k}, B_{1:k-1}),~~~~\phi_{2} =
\mathrm{tri}(A_{k+1:N_{y}}, B_{k+1:N_{y}-1}),~~~~\mathrm{and}
\\
\\
X &= \begin{pmatrix}
 0&\cdots&-B_k^T&0&\cdots&0\\
 0&\cdots&0&-B_k&\cdots&0
 \end{pmatrix}^T,~ 
Y = \begin{pmatrix}
 0&\cdots&0&I&\cdots&0\\
 0&\cdots&I&0&\cdots&0
\nonumber 
 \end{pmatrix}.
\end{align*}

\noindent Thus, the original block tridiagonal matrix can be decomposed into the sum of a 
block diagonal matrix (with its two diagonal blocks themselves being block
tridiagonal) and a correction term parametrized by the $N_{x} \times N_{x}$
matrix $B_{k}$, which we will refer to as the ``bridge matrix''.
Using the matrix inversion lemma, we have
\begin{equation*}
G^R=(\tilde{K} + XY)^{-1} 
=\tilde{K}^{-1}-(\tilde{K}^{-1}X)
\left(I + Y\tilde{K}^{-1}X\right)^{-1}(Y\tilde{K}^{-1}),
\end{equation*}
where
\begin{eqnarray}
\label{eqn:minv}
\tilde{K}^{-1}X & = & 
\begin{pmatrix}
-\phi_{1}^{-1}\left(:,k\right)B_{k}
&0\\
0&
-\phi_{2}^{-1}\left(:,1\right)B_{k}^{T}
\end{pmatrix},
\\
\left(I + Y\tilde{K}^{-1}X\right)^{-1} & = &
\begin{pmatrix}
I&-\phi_{2}^{-1}\left(1,1\right)B_{k}^{T}\\
-\phi_{1}^{-1}\left(k,k\right)B_{k}&I
\end{pmatrix}^{-1},
\nonumber \\
Y\tilde{K}^{-1} & = &
\begin{pmatrix}
0&\phi_{2}^{-1}\left(:,1\right)^{T} \\
\phi_{1}^{-1}\left(:,k\right)^{T}
&0
\end{pmatrix},
\nonumber 
\end{eqnarray}
and $\phi_{1}^{-1}\left(:,k\right)$ and
$\phi_{2}^{-1}\left(:,1\right)$ 
denote respectively the last and first block columns of $\phi_{1}^{-1}$ and $\phi_{2}^{-1}$.

This shows that the entries of $\tilde{K}^{-1}$ are modified through the entries
 from the first rows and last columns of $\phi_1^{-1}$ and $\phi_2^{-1}$, 
as well as the bridge matrix $B_k$.
Specifically, since $\phi_1$ is before or ``above'' the bridge
point we only need the last column of its inverse to 
reconstruct $G^R$.  Similarly, 
since $\phi_2$ is after or ``below'' the bridge
point we only need the first column of its inverse.
These observations were noted in~\cite{cauley:07},
where the authors demonstrated a parallel divide-and-conquer
 approach to determine the diagonal entries for the inverse of block 
tridiagonal matrices.
We begin by generalizing the method from~\cite{cauley:07} in
 order to compute all information necessary to 
determine the distributed compact representation of $G^R$~(\ref{eqn:uv}).
That is, we would like to create a combining methodology 
for sub-matrix inverses with two major goals in mind.
 First, it must allow for the calculation of all information that would
be required to repeatedly join sub-matrix inverses, in order to mimic
 the combining process shown in Figure~\ref{fig:decomp_over}.
Second, at the final stage of the combining process
it must facilitate the computation of the block tridiagonal portion 
for the combined inverses.
Details pertaining to the parallel computation of $G^R$ are 
provided in Appendix~\ref{ap:GR}.
 The time complexity of the algorithm presented
is $O({N_{x}^{3}N_{y}}/{p} + N_{x}^{3}\log p)$, with memory consumption
$O\left({N_{x}^{2}N_{y}}/{p} + N_{x}^{2}\right)$.  
The distribution of the compact representation
is at the foundation of an efficient parallel method for calculating the
less-than Green's Function $G^{<}$ and greater-than Green's Function $G^{>}$.

\section{Parallel Computation of the Less-than Green's Function}
\label{s:pmult}

The parallel inversion algorithm described 
in Section~\ref{s:pinv_blktri} not only has advantages in computational and 
memory efficiency but also facilitates the formulation of a fast, and highly scalable, parallel 
matrix multiplication algorithm.  This plays an important role during the simulation process 
due to the fact that computation of the less-than Green's Function requires 
matrix products with the retarded Green's Function matrix:

\begin{align}
G^{<} = G^{R} \Sigma^{<} {G^{R}}^{*}.
\label{eqn:gless_main}
\end{align}

\noindent 
$\Sigma^{<}$, which we will refer to as the less-than 
scattering matrix, is typically assumed to be a block 
diagonal matrix.
We will demonstrate how the distributed compact representation 
of the semiseparable matrix $G^{R}$ presented in Section~\ref{s:pinv_blktri} 
can be used to calculate the necessary information from $G^<$.
Specifically, the electron density for the device will be
calculated through the diagonal entries of $G^{<}$ and the current 
characteristics through the first off-diagonal blocks of $G^{<}$.

\subsection{Mathematical Description}
\label{s:glessmath}
Recall that our initial state for this procedure would assume that
portions of the block tridiaongal (corresponding to the size and location of the
 divisions) of $G^{R}$ have been calculated and stored.
It is important to note that there are many generator representations for 
$G^{R}$ and we would like to select a representation that will facilitate 
efficient calculation of $G^{<}$.
For the mathematical operation shown in~(\ref{eqn:gless_main}),
our starting point will be describing the $k^{\mathrm{th}}$ block row of $G^{R}$ 
in terms of $D_{i},~\left(g^{\overrightarrow{R}}_{i}\right)^T,~\mathrm{and}~g^{\underleftarrow{R}}_{i}$, the diagonal blocks and 
two generator sequences respectively.
 The following expressions are used to determine
the generators from the block tridiagonal of the semiseparable matrix:

\begin{align}
\label{eqn:ratio_back}
P_{i} &= g^{\underleftarrow{R}}_{i} D_{i+1} \Rightarrow g^{\underleftarrow{R}}_{i} = P_{i} \left( D_{i+1} \right)^{-1}, \nonumber \\ 
\nonumber \\
Q_{i} &= \left(g^{\overrightarrow{R}}_{i}\right)^{T} D_{i} \Rightarrow \left(g^{\overrightarrow{R}}_{i}\right)^{T} = Q_{i} \left( D_{i} \right)^{-1}. 
\end{align}

\noindent 
$D_{i},~P_{i},~\mathrm{and}~Q_{i}$, are the diagonal, upper diagonal, 
and lower diagonal blocks respectively.
Thus, the generators $\left(g^{\overrightarrow{R}}_i\right)^T$ and $g^{\underleftarrow{R}}_{i}$ are used to describe the
$k^{\mathrm{th}}$ block row of $G^{R}$ 
semiseparable matrix in the following way: 

\begin{align*}
&G^{R} \left(k, : \right) = \\
&~~\begin{pmatrix}
\prod\limits_{i=k-1}^{1}\left(g^{\overrightarrow{R}}_{i}\right)^{T} D_{1} &
\cdots &
\left(g^{\overrightarrow{R}}_{k-1}\right)^{T} D_{k-1} &
D_{k} &
g^{\underleftarrow{R}}_{k} D_{k+1} &
\cdots
\prod\limits_{i=k}^{N_{y}-1}g^{\underleftarrow{R}}_{i} D_{N_{y}} &
\end{pmatrix}
\end{align*}

\noindent This generator representation for $G^{R}$
along with the block diagonal structure of $\Sigma^{<}$ 
allows for us to express each diagonal block of $G^{<}$
in terms of recursive sequences.
Both the forward recursive sequence $g^{\overrightarrow{<}}_{i}$ and backward recursive 
sequence $g^{\underleftarrow{<}}_{i}$ are dependent on a common sequence of injections terms
$J_{i}$ (note: the arrow orientation for $g^{<}$ matches that of $g^{R}$).
  The relationships between the sequences and the diagonal blocks
of $G^{<}$ are shown below:

\begin{align}
\label{eqn:gless_rec}
&J_{i} = D_{i} \Sigma^{<}_{i} D_{i}^{*},~\quad i = 1, 2, \ldots, N_{y} \nonumber \\
\nonumber
\\
\nonumber
&g^{\overrightarrow{<}}_{1} = J_{1}\\
\nonumber
&g^{\overrightarrow{<}}_{i} = J_{i} + \left(g^{\overrightarrow{R}}_{i-1}\right)^{T} g^{\overrightarrow{<}}_{i-1} \left(g^{\overrightarrow{R}}_{i-1}\right)^{C},~\quad i = 2, \ldots, N_{y} \\
\nonumber\\
&g^{\underleftarrow{<}}_{N_{y}} = J_{N_{y}}\\
\nonumber
&g^{\underleftarrow{<}}_{i} = J_{i} + g^{\underleftarrow{R}}_{i} g^{\underleftarrow{<}}_{i+1} \left(g^{\underleftarrow{R}}_{i}\right)^{*},~\quad i = N_{y}-1, \ldots, 1 \\
\nonumber
\\
\nonumber
&G^{<}(1, 1) = g^{\underleftarrow{<}}_{1}\\
\nonumber
&G^{<}(i, i) = g^{\underleftarrow{<}}_{i} + g^{\overrightarrow{<}}_{i} - J_{i},~\quad i = 2, \ldots, N_{y} 
\end{align}

\noindent Similar serial recursions have 
been shown in~\cite{anant} and~\cite{jjain:tr07}.
Our strategy in this work is to exploit the distributed compact representation 
of $G^R$ in order to produce these sequences efficiently.
That is, we will divide the computation needed to calculate the three sequences 
$J_{i}$, $g^{\overrightarrow{<}}_{i}$, and $g^{\underleftarrow{<}}_{i}$.
into several sub-problems that can
be efficiently separated across many processors.

As motivation, if we assume that there are $N_y = 2N$ blocks, 
we can define two sub-problems by separating 
$\Sigma^{<} = \Sigma^{<;1} + \Sigma^{<;2}$, where $\Sigma^{<;1}$ contains blocks 
$1$ through $N$ and $\Sigma^{<;2}$ contains blocks $N+1$ through $2N$.
Then, we can write: 

\begin{align*}
G^{<} = G^{r} \Sigma^{<} {G^{r}}^{*} = 
G^{r} \left( \Sigma^{<;1} + \Sigma^{<;2}\right) {G^{r}}^{*}
=G^{r}  \Sigma^{<;1} {G^{r}}^{*} +
G^{r}  \Sigma^{<;2} {G^{r}}^{*} 
=G^{<;1} + G^{<;2}.
\end{align*}

\noindent For this example, we have assumed an equal separation of the 
diagonal blocks for the scattering matrix
 $\Sigma^{<}$, i.e. $\Sigma^{<;1}_{i} = 0,~~\forall i > N$.  Thus, 
for the first sub-problem we have:

\begin{align*}
&J_{i;1} = D_{i} \Sigma^{<;1}_{i} D_{i}^{*},~\quad &i = 1, 2, \ldots, N \\
&J_{i;1} = 0 ,~\quad &i = N+1, 2, \ldots, 2N \\
\\
&g^{\overrightarrow{<}}_{1;1} = J_{1;1}\\
&g^{\overrightarrow{<}}_{i;1} = J_{i;1} + \left(g^{\overrightarrow{R}}_{i-1}\right)^{T} g^{\overrightarrow{<}}_{i-1;1} \left(g^{\overrightarrow{R}}_{i-1}\right)^{C},~\quad &i = 2, \ldots, N \\
&g^{\overrightarrow{<}}_{i;1} = \left(g^{\overrightarrow{R}}_{i-1}\right)^{T} g^{\overrightarrow{<}}_{i-1;1} \left(g^{\overrightarrow{R}}_{i-1}\right)^{C},~\quad &i = N+1, \ldots, 2N \\
\\
&g^{\underleftarrow{<}}_{i;1} = 0,~\quad &i = 2N, \ldots, N+1 \\
&g^{\underleftarrow{<}}_{N;1} = J_{N;1}\\
&g^{\underleftarrow{<}}_{i;1} = J_{i;1} + g^{\underleftarrow{R}}_{i} g^{\underleftarrow{<}}_{i+1;1} \left(g^{\underleftarrow{R}}_{i}\right)^{*},~\quad &i = N-2, \ldots, 1 \\
\end{align*}

\noindent We then see the following relationships for the second sub-problem:

\begin{align*}
&J_{i;2} = 0 ,~\quad &i = 1, 2, \ldots, N \\
&J_{i;2} = D_{i} \Sigma^{<;2}_{i} D_{i}^{*},~\quad &i = N+1, 2, \ldots, 2N \\
\\
&g^{\overrightarrow{<}}_{i;2} = 0,~\quad &i = 1, \ldots, N \\
&g^{\overrightarrow{<}}_{N+1;2} = J_{N+1;2}\\
&g^{\overrightarrow{<}}_{i;2} = J_{i;2} + \left(g^{\overrightarrow{R}}_{i-1}\right)^{T} g^{\overrightarrow{<}}_{i-1;2} \left(g^{\overrightarrow{R}}_{i-1}\right)^{C},~\quad &i = N+2, \ldots, 2N \\
\\
&g^{\underleftarrow{<}}_{2N;2} = J_{2N;2}\\
&g^{\underleftarrow{<}}_{i;2} = J_{i;2} + g^{\underleftarrow{R}}_{i} g^{\underleftarrow{<}}_{i+1;2} \left(g^{\underleftarrow{R}}_{i}\right)^{*},~\quad &i = 2N-1, \ldots, N+1 \\
&g^{\underleftarrow{<}}_{i;2} = g^{\underleftarrow{R}}_{i} g^{\underleftarrow{<}}_{i+1;2} \left(g^{\underleftarrow{R}}_{i}\right)^{*},~\quad &i = N, \ldots, 1 \\
\end{align*}

\begin{figure}
  \centering
  \subfigure[
  The non-zero matrices that need to be computed for sub-problem 1.
  The terms that are computed on CPU 1 are shaded.    
  ]{\label{fig:gless_2cpu_1}\includegraphics[width=0.5\textwidth]{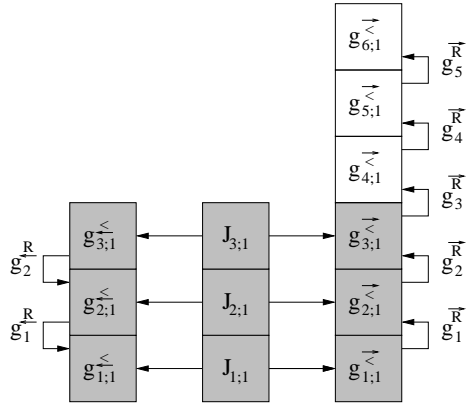}}                
  \quad\subfigure[
  The non-zero matrices that need to be computed for sub-problem 2.
  The terms that are computed on CPU 2 are shaded.    
  ]{\label{fig:gless_2cpu_2}\includegraphics[width=0.5\textwidth]{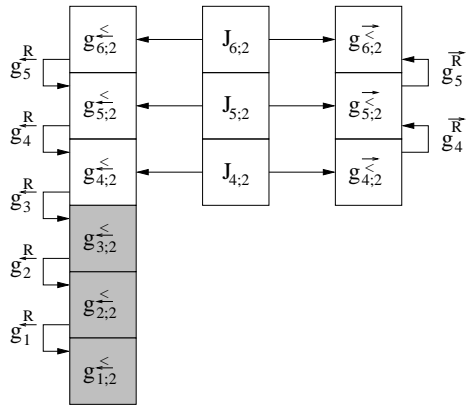}}
  \caption{Distribution of $G^{<}$ computation into two sub-problems across two CPUs.}
  \label{fig:gless_2cpu}
\end{figure}

\noindent The recursions for the case of two sub-problems 
are illustrated in Figure~\ref{fig:gless_2cpu} with $N_y = 2N = 6$.
  Here, the shaded terms in 
Figure~\ref{fig:gless_2cpu_1} and Figure~\ref{fig:gless_2cpu_2} represent the terms for
each sub-problem that will be computed on CPU 1 and CPU 2 respectively.
In summary, the separation of the diagonal blocks, the generator sequences, and
the scattering matrix evenly across two computers will result in the following 
order of operations:

\begin{align*}
&\mathrm{CPU~1}&\mathrm{CPU~2}&\\
\\
\mathrm{Stage~I}~\quad~\quad&J_{i;1} = D_{i} \Sigma^{<;1}_{i} D_{i}^{*}, &
J_{i;2} = D_{i} \Sigma^{<;2}_{i} D_{i}^{*}, \\
&i = 1, 2, \ldots, N &
i = N+1, 2, \ldots, 2N \\
\\
\mathrm{Stage~II}~\quad~\quad&g^{\overrightarrow{<}}_{1;1} = J_{1;1}&
g^{\underleftarrow{<}}_{2N;2} = J_{2N;2} \\
&g^{\overrightarrow{<}}_{i;1} = J_{i;1} + \left(g^{\overrightarrow{R}}_{i-1}\right)^{T} g^{\overrightarrow{<}}_{i-1;1} \left(g^{\overrightarrow{R}}_{i-1}\right)^{C},&
g^{\underleftarrow{<}}_{i;2} = J_{i;2} + g^{\underleftarrow{R}}_{i} g^{\underleftarrow{<}}_{i+1;2} \left(g^{\underleftarrow{R}}_{i}\right)^{*}, \\
&i = 2, \ldots, N&
i = 2N-1, \ldots, N+1 \\
\\
\mathrm{Stage~III}~\quad~\quad&g^{\underleftarrow{<}}_{i;2} = g^{\underleftarrow{R}}_{i} g^{\underleftarrow{<}}_{i+1;2} \left(g^{\underleftarrow{R}}_{i}\right)^{*},&
g^{\overrightarrow{<}}_{i;1} = \left(g^{\overrightarrow{R}}_{i-1}\right)^{T} g^{\overrightarrow{<}}_{i-1;1} \left(g^{\overrightarrow{R}}_{i-1}\right)^{C},\\
&i = N, \ldots, 1 &
i = N+1, \ldots, 2N \\
\\
\mathrm{Stage~IV}~\quad~\quad&g^{\underleftarrow{<}}_{N;1} = J_{N;1} &
g^{\overrightarrow{<}}_{N+1;2} = J_{N+1;2} \\
&g^{\underleftarrow{<}}_{i;1} = J_{i;1} + g^{\underleftarrow{R}}_{i} g^{\underleftarrow{<}}_{i+1;1} \left(g^{\underleftarrow{R}}_{i}\right)^{*}, &
g^{\overrightarrow{<}}_{i;2} = J_{i;2} + \left(g^{\overrightarrow{R}}_{i-1}\right)^{T} g^{\overrightarrow{<}}_{i-1;2} \left(g^{\overrightarrow{R}}_{i-1}\right)^{C}, \\
&i = N-2, \ldots, 1&
i = N+2, \ldots, 2N \\
\end{align*}

\noindent If we consider multiplication to be 
of order $N_x^3$, then on a single processor the 
$G^{<}$ calculation would require 
$3 \times (2N_y) N_x^3 =6 N_yN_x^3$ operations.
In this case we have four stages, each with 
$2 \times \frac{N_y}{2} N_x^3= N_yN_x^3$ operations.  Therefore,
using two processors we have reduced the number
of multiplications to $4 N_yN_x^3$.
We now generalize this process and demonstrate further 
speed-up as both the number of processors and the length of 
the device increase.

\subsection{Parallel Implementation}
\label{s:glessparallel}

In order to simplify the presentation of the method, 
we will assume that the number of sub-problems $p$ 
evenly divides the total number of blocks $N_y = pN$.
In general this assumption is not required.
We will separate the $G^{<}$ operation 
evenly into $p$ sub-problems by dividing 
$\Sigma^{<} = \Sigma^{<;1} + \Sigma^{<;2} + \ldots \Sigma^{<;p}$,
 where $\Sigma^{<;k}$ contains the $k^{\mathrm{th}}$ portion 
of the less-than scattering matrix.
Given that $\Sigma^{<;k}_{i} = 0,~~\forall i > kN~\mathrm{and}~i < (k-1)N + 1$,
the $k^{\mathrm{th}}$ sub-problem will be solved through the following recursions:

\begin{align*}
&J_{i;k} = 0 ,~\quad &i = 1, \ldots, (k-1)N \\
&J_{i;k} = D_{i} \Sigma^{<;k}_{i} D_{i}^{*},~\quad &i = (k-1)N+1, \ldots, kN \\
&J_{i;k} = 0 ,~\quad &i = kN+1, \ldots, pN \\
\\
&g^{\overrightarrow{<}}_{i;k} = 0,~\quad &i = 1, \ldots, (k-1)N \\
&g^{\overrightarrow{<}}_{i;k} = J_{1;k},~\quad &i = (k-1)N + 1\\
&g^{\overrightarrow{<}}_{i;k} = J_{i;k} + \left(g^{\overrightarrow{R}}_{i-1}\right)^{T} g^{\overrightarrow{<}}_{i-1;k} \left(g^{\overrightarrow{R}}_{i-1}\right)^{C},~\quad &i = (k-1)N+2, \ldots, kN \\
&g^{\overrightarrow{<}}_{i;k} = \left(g^{\overrightarrow{R}}_{i-1}\right)^{T} g^{\overrightarrow{<}}_{i-1;k} \left(g^{\overrightarrow{R}}_{i-1}\right)^{C},~\quad &i = kN+1, \ldots, pN \\
\\
&g^{\underleftarrow{<}}_{i;k} = 0,~\quad &i = pN, \ldots, kN+1 \\
&g^{\underleftarrow{<}}_{i;k} = J_{i;k},~\quad &i = kN \\
&g^{\underleftarrow{<}}_{i;k} = J_{i;k} + g^{\underleftarrow{R}}_{i} g^{\underleftarrow{<}}_{i+1;k} \left(g^{\underleftarrow{R}}_{i}\right)^{*},~\quad &i = kN-1, \ldots, (k-1)N+1 \\
&g^{\underleftarrow{<}}_{i;k} = g^{\underleftarrow{R}}_{i} g^{\underleftarrow{<}}_{i+1;k} \left(g^{\underleftarrow{R}}_{i}\right)^{*},~\quad &i = (k-1)N, \ldots, 1 \\
\end{align*}

\noindent So, in a sense by splitting the scattering matrix we have divided the
injection sequence $J$ evenly and created two new sub-problem propagating sequences $g^{\overrightarrow{<}}_{i;k}$ 
and $g^{\underleftarrow{<}}_{i;k}$.  
However, the sub-problem propagating sequences have the special property that they 
do not start (are zero) until they reach the indices governed by the sub-problem.  
This is due to the fact that there are no injection terms for the given sub-problem
until these points are reached.
In addition, once outside the range of the sub-problem we do not have any additional 
 injection terms in the recursions.  Thus, we have a standard two-sided auto-regressive 
expression where the terms are simply propagated by multiplication with generator matrices.

\begin{figure}
  \centering
  \subfigure[
  The non-zero matrices that need to be computed for sub-problem 1.
  The terms that are computed on CPU 1 are shaded.    
  ]{\label{fig:gless_4cpu_1}\includegraphics[width=0.5\textwidth]{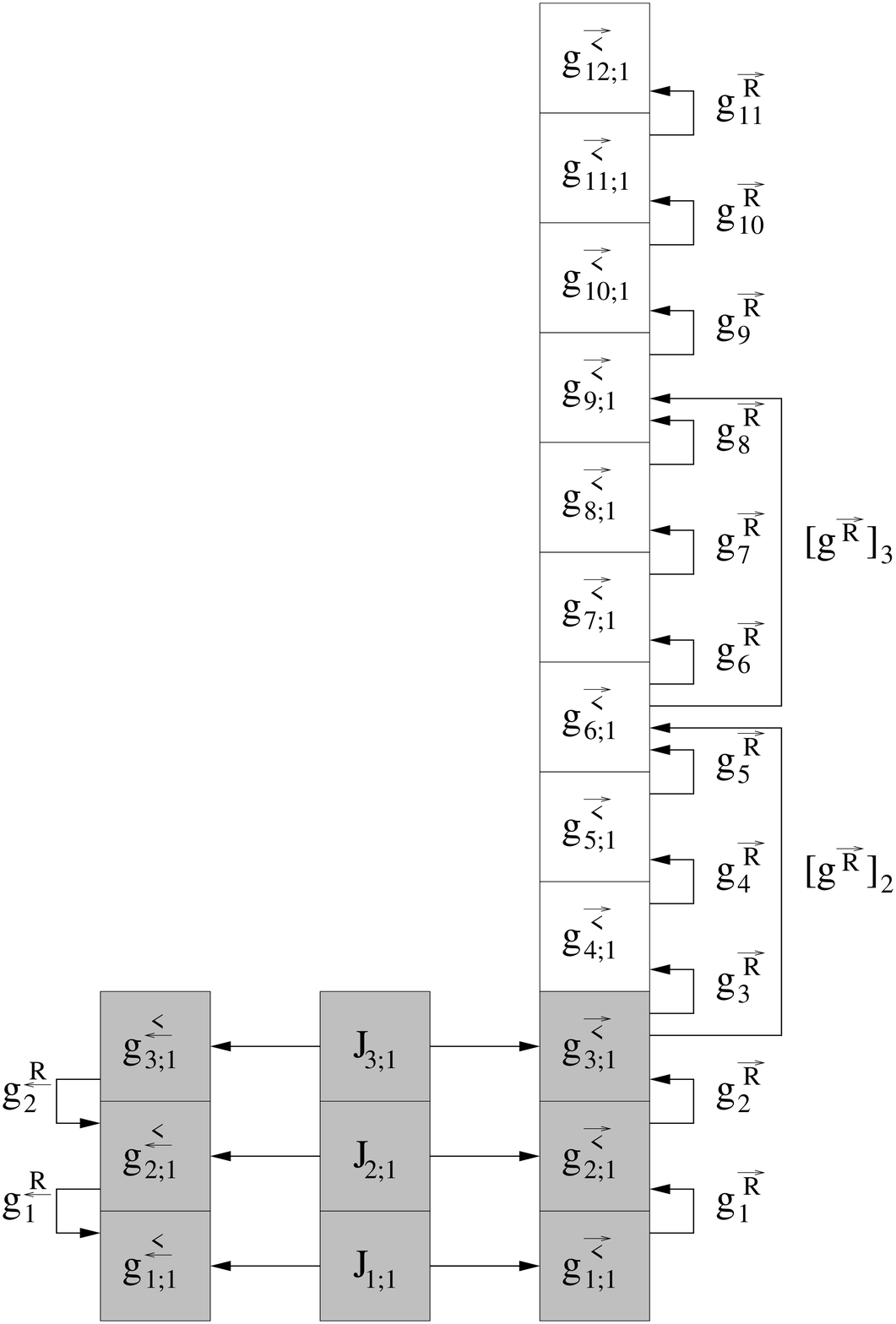}}                
  \quad\subfigure[
  The non-zero matrices that need to be computed for sub-problem 4.
  The terms that are computed on CPU 4 are shaded.    
  ]{\label{fig:gless_4cpu_4}\includegraphics[width=0.5\textwidth]{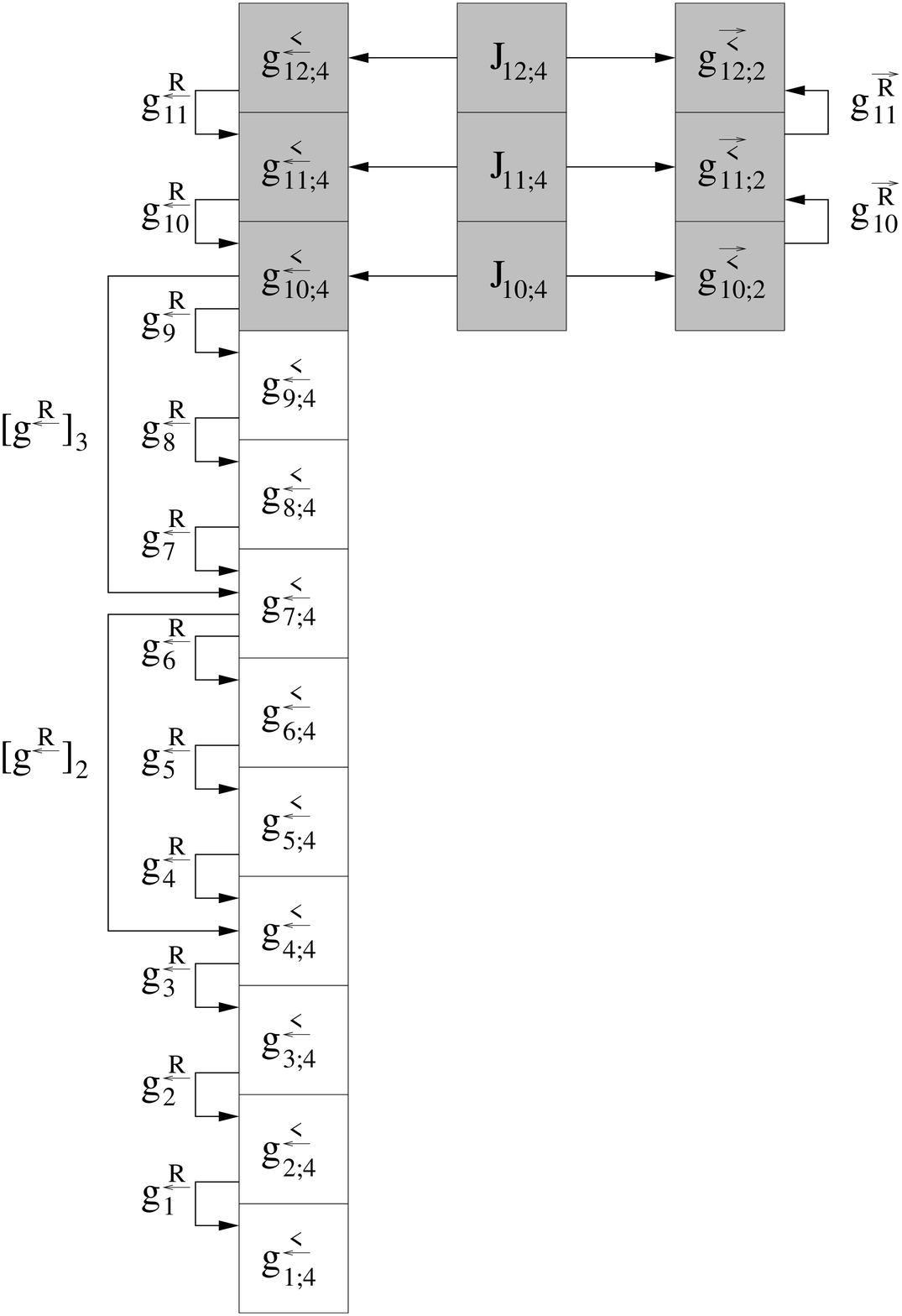}}
  \caption{Distribution of $G^{<}$ computation into four sub-problems across four CPUs.}
  \label{fig:gless_4cpu}
\end{figure}

The recursions for the case of four sub-problems are illustrated in
 Figure~\ref{fig:gless_4cpu} for an example with $N_y = 4N = 12$.  Here, the shaded terms in 
Figure~\ref{fig:gless_4cpu_1} and Figure~\ref{fig:gless_4cpu_4} represent the terms for
each sub-problem that are computed on CPU 1 and CPU 4 respectively.
For each sub-problem we introduce two new sequences that will be referred to as
 ''skip matrices''.  Specifically, we define for each processor $k$ two matrices 
$\left[g^{\underleftarrow{R}}\right]_k$ and $\left[g^{\overrightarrow{R}}\right]_k^T$ that are accumulations of the generator 
terms stored on the processor.  As can be clearly seen from Figure~\ref{fig:gless_4cpu}
these skip matrices allow for several steps in the recursive process to be preformed 
through a single operation.
Therefore, if the skip matrices are made available to each processor several terms
for each sub-problem can be found concurrently.
In Figure~\ref{fig:gless_4cpu_4}, we can see that skip matrices will allow for
$g^{\underleftarrow{<}}_{9;4}$, $g^{\underleftarrow{<}}_{6;4}$, and $g^{\underleftarrow{<}}_{3;4}$ to be determined currently 
by CPUs 3, 2, and 1 respectively.

Given that each sub-problem will produce both a forward and backward propagating 
sequence it should be clear that CPU $k$ will need to preform computation 
for $g^{\overrightarrow{<}}$ sequences from sub-problems $\{ k+1, k+2, \ldots, p\}$ 
and $g^{\underleftarrow{<}}$ sequences from sub-problems $\{ 1,2, \ldots, k-1\}$.
However, each $g^{\overrightarrow{<}}$ sequence and $g^{\underleftarrow{<}}$ sequence from other sub-problems 
may be combined before the generators on that CPU are applied.  
This is due to the fact that the sequences from each sub-problem will 
eventually be added together to form $G^{<} = \sum\limits_{k = 1}^{p} G^{<;k}$.
For the example shown in Figure~\ref{fig:gless_4cpu} we see that
 CPU $1$ will first need to form: 
\begin{align*}
g^{\underleftarrow{<}}_{4;4} + g^{\underleftarrow{<}}_{4;3} + g^{\underleftarrow{<}}_{4;2} &= 
\left[g^{\underleftarrow{R}}\right]_2 
\left(\left[g^{\underleftarrow{R}}\right]_3 g^{\underleftarrow{<}}_{10;4} \left[g^{\underleftarrow{R}}\right]_3^* 
+ g^{\underleftarrow{<}}_{7;3} \right) \left[g^{\underleftarrow{R}}\right]_2^* + g^{\underleftarrow{<}}_{4;2},
\end{align*}

\noindent before the required terms for each sub-problem can be
calculated.
That is, after the lead term for the backward propagating sequence:
$\left( g^{\underleftarrow{<}}_{4;4} + g^{\underleftarrow{<}}_{4;3} + g^{\underleftarrow{<}}_{4;2} \right)$ has been computed, 
the generators governed by CPU 1: $g^{\underleftarrow{R}}_{3}$, $g^{\underleftarrow{R}}_{2}$, and $g^{\underleftarrow{R}}_{1}$,
can be applied to fulfill the sub-problem solutions.
We now have all the tools necessary to construct a general procedure:

\hspace{20 mm}
\begin{enumerate}
\item Compute the skip products of each generator sequence $g^{\overrightarrow{R}}$ and $g^{\underleftarrow{R}}$, i.e.
compute $\left[g^{\underleftarrow{R}} \right]_k= \prod\limits_{i = (k-1)N+1 }^{kN} {g^{\underleftarrow{R}}_{i}}$ 
and $\left[g^{\overrightarrow{R}} \right]_k^T = \prod\limits_{i = kN-1}^{(k-1)N} {\left(g^{\overrightarrow{R}}_{i}\right)^{T}}$,
where we will define $\left(g^{\overrightarrow{R}}_0\right)^T = g^{\underleftarrow{R}}_{N_y} = I$.   
\item Compute the initial injection and propagation terms for each sub-problem: 
\begin{align*}
&J_{i;k} = D_{i} \Sigma^{<;k}_{i} D_{i}^{*},~\quad &i = (k-1)N+1, \ldots, kN \\
\\
&g^{\overrightarrow{<}}_{i;k} = J_{i;k},~\quad &i = (k-1)N + 1\\
&g^{\overrightarrow{<}}_{i;k} = J_{i;k} + \left(g^{\overrightarrow{R}}_{i-1}\right)^{T} g^{\overrightarrow{<}}_{i-1;k} \left(g^{\overrightarrow{R}}_{i-1}\right)^{C},~\quad &i = (k-1)N+2, \ldots, kN \\
\\
&g^{\underleftarrow{<}}_{i;k} = J_{i;k},~\quad &i = kN \\
&g^{\underleftarrow{<}}_{i;k} = J_{i;k} + g^{\underleftarrow{R}}_{i} g^{\underleftarrow{<}}_{i+1;k} \left(g^{\underleftarrow{R}}_{i}\right)^{*},~\quad &i = kN-1, \ldots, (k-1)N+1 \\
\end{align*}
\item Transfer forward and backward skip products: 
$\left[g^{\overrightarrow{R}} \right]_k^T$ and $\left[g^{\underleftarrow{R}} \right]_k$, 
as well as forward and backward lead propagating matrices: 
$g^{\overrightarrow{<}}_{kN;k}$ and $g^{\underleftarrow{<}}_{(k-1)N+1;k}$,
for each sub-problem $k$ to all CPUs.  This will be a total of $4pNx^2$ entries.

\item CPU $k$ will 
construct combined lead propagating term for $g^{\underleftarrow{<}}$ sequences from 
sub-problems $\{ k+1, k+2, \ldots, p\}$.
\begin{align*}
\sum\limits_{j=k+1}^{p} g^{\underleftarrow{<}}_{kN+1;j} &= 
g^{\underleftarrow{<}}_{kN+1;k+1} +
\sum\limits_{j=k+2}^{p} 
 \left(
\prod\limits_{l=k+1}^{j-1} \left[g^{\underleftarrow{R}} \right]_l 
\right)
 g^{\underleftarrow{<}}_{(j-1)N+1;j} 
\left(
\prod\limits_{l=j-1}^{k+1} \left[g^{\underleftarrow{R}} \right]_l^* 
\right)
\end{align*}
\noindent CPU $k$ will 
construct combined lead propagating term for $g^{\overrightarrow{<}}$ sequences from 
sub-problems $\{ 1,2, \ldots, k-1\}$.
\begin{align*}
\sum\limits_{j=1}^{k-1} g^{\overrightarrow{<}}_{(k-1)N;j} &= 
g^{\overrightarrow{<}}_{(k-1)N;k-1} +
\sum\limits_{j=1}^{k-2} 
 \left(
\prod\limits_{l=k-1}^{j+1} \left[g^{\overrightarrow{R}} \right]_l^T 
\right)
 g^{\overrightarrow{<}}_{jN+1;j} 
\left(
\prod\limits_{l=j+1}^{k-1} \left[g^{\overrightarrow{R}} \right]_l^C 
\right)
\end{align*}
\item Successively multiply by the governed
 generator terms $g^{\underleftarrow{R}}$ and $g^{\overrightarrow{R}}$ in order fulfill sub-problem solutions.
\begin{align*}
\sum\limits_{j=k+1}^{p} g^{\underleftarrow{<}}_{i;j} &= 
 \left(
\prod\limits_{l=i}^{kN} g^{\underleftarrow{R}}_l 
\right)
\sum\limits_{j=k+1}^{p} g^{\underleftarrow{<}}_{kN+1;j} 
\left(
\prod\limits_{l=kN}^{i} \left(g^{\underleftarrow{R}}_l\right)^* 
\right),
\\
\quad&i = kN, \ldots, (k-1)N+1. 
\\
\\
\sum\limits_{j=1}^{k-1} g^{\overrightarrow{<}}_{i;j} &= 
 \left(
\prod\limits_{l=(i-1)}^{(k-1)N} \left(g^{\overrightarrow{R}}_l\right)^T 
\right)
\sum\limits_{j=1}^{k-1} g^{\overrightarrow{<}}_{(k-1)N;j} 
\left(
\prod\limits_{l=(k-1)N}^{(i-1)} \left(g^{\overrightarrow{R}}_l\right)^C 
\right),
\\
\quad&i = (k-1)N+1, \ldots, kN.
\end{align*}
\item Each CPU will combine portions of all sub-problem solutions in order
to produce corresponding portion of the diagonal blocks of $G^{<}$
based upon the relationships shown in~(\ref{eqn:gless_rec}). 
\end{enumerate}
\hspace{20 mm}

In order to analyze the computational 
improvement for the approach we consider
the number of multiplications required for each stage.
The accumulation of the skip matrices described in Step 1
will result in $2 N_x^3 N_y / p$ multiplications.
The individual sub-problem recursions of Step 2 results
in $6 N_x^3 N_y / p$ multiplications.  
Step 4 describes how the skip matrices may be applied in order 
to create the sum for each sub-problem propagating sequence, 
requiring $2 N_x^3 (p - 2)$ multiplications.
Finally, these two sums must be propagated through each governed 
generator matrices, requiring $4 N_x^3 N_y / p$ multiplications.
If one is interested in also computing the
current characteristics for the device, they may be determined 
through the off-diagonal blocks of $G^{<}$:

\begin{align}
&G^{<}(i, i+1) = g^{\underleftarrow{<}}_{i+1} \left(g^{\underleftarrow{R}}_i\right)^* + \left(g^{\overrightarrow{R}}_i\right)^T g^{\overrightarrow{<}}_{i},~\quad i = 1, \ldots, N_{y}-1. 
\end{align}

\noindent As portions of each generator sequence and propagating sequence have
been evenly distributed amongst the processors, the resulting computation 
would be $2 N_x^3 N_y / p$.
Therefore, the total number of multiplications for the approach 
is $2 N_x^3 (p - 2) + 14 N_x^3 N_y / p$ compared to $6 N_x^3 N_y$
for the serial algorithm.  Therefore, the speedup of our approach 
is 
\begin{align}
\label{eqn:gless_speed}
\frac{p}{2.34 + \frac{p(p-2)}{3N_y}}
\end{align}

\noindent  We can clearly see that
if $N_y >> p$ we will approach a speedup of $p / 2.34$.

\begin{table}
\footnotesize
\begin{center}
\begin{tabular}
{|c||c|c|c|c||c|c|c|c||c|c|c|c|c|c|c|c|c|}
\hline
& \multicolumn{4}{|c||}{$2$nm cross-section} 
& \multicolumn{4}{|c||}{$3$nm cross-section} 
& \multicolumn{4}{|c|}{$4$nm cross-section} \\ 
\hline
$p$&4&8&16&32&4&8&16&32&4&8&16&32 \\
\hline
\hline
& \multicolumn{12}{|c|}{$35$nm length} \\
\hline
$G^{R}$&
35.8&21.5&16.2&N/A&411.1&241.4&163.6&N/A&1915.5&1099.7&726.1& N/A \\
\hline
$G^{<}$&
22.7&12.7&10.0&N/A&253.7&133.7&81.2&N/A&1159.2&598.5&349.7& N/A\\
\hline
RGF$\times$ &
1.5&2.7&3.4&N/A&1.4&2.6&4.1&N/A&1.4&2.6&4.2& N/A\\
\hline
\hline
& \multicolumn{12}{|c|}{$53$nm length} \\
\hline
$G^{R}$&
51.4&29.3&19.7&N/A&594.4&331.9&211.4&N/A&2756.1&1523.4&946.4& N/A\\
\hline
$G^{<}$&
33.5&18.5&13.8&N/A&380.9&198.5&113.8&N/A&1741.6&887.8&489.1& N/A\\
\hline
RGF$\times$ &
1.6&2.8&4.0&N/A&1.5&2.7&4.5&N/A&1.4&2.7&4.6& N/A\\
\hline
\hline
& \multicolumn{12}{|c|}{$70$nm length} \\
\hline
$G^{R}$&
67.1&38.8&23.8&17.7&779.0&422.6&255.0&181.3&3601.3&1953.7&1154.5&782.1 \\
\hline
$G^{<}$&
44.6&24.4&15.8&12.9&507.6&261.5&145.5&103.4&2324.2&1182.2&640.2&408.9 \\
\hline
RGF$\times$ &
1.6&2.8&4.5&5.7&1.5&2.8&4.8&6.8&1.4&2.7&4.9&7.4 \\
\hline
\end{tabular}
\end{center}
\caption{Runtime comparisons between parallel $G^{R}$ and $G^{<}$ algorithms
and serial RGF approach.  Silicon nanowires with lengths $35$nm, $53$nm,
and $70$nm are examined with cross-sections of $2$nm, $3$nm, and $4$nm.
''RGF$\times$'' is the observed speed-up for the combined time.
''N/A'' is used for devices that are too short
to support the number of CPUs.}
\label{tab:gless_scale}
\end{table}

\section{Results}
\label{ap:results}

The parallel $G^{R}$ and $G^{<}$ algorithms have been implemented
in C using MPI for inter-processor communication.
All computational complexity analyses
were performed on a cluster of 
Intel E5410 processors with $16$GB of shared memory 
for the $8$ core machines.
The OMEN simulator~\cite{mLU_08} was used to 
perform simulations for square silicon nanowires 
employing the $sp3d5s*$ atomistic tight-binding model with electron-phonon
scattering.
We will begin by demonstrating the computational 
efficiency of our approach.  We will then analyze
device characteristics for large cross-section
nanowires that have prohibitive memory
requirements for the serial RGF approach.

\begin{figure}
  \centering
  \subfigure[
  $G^{<}$ speed-up is compared against theoretical estimate
from~(\ref{eqn:gless_speed}) for $50$nm length silicon nanowires.
  ]{\label{fig:gless_scale_50nm}\includegraphics[width=0.8\textwidth]{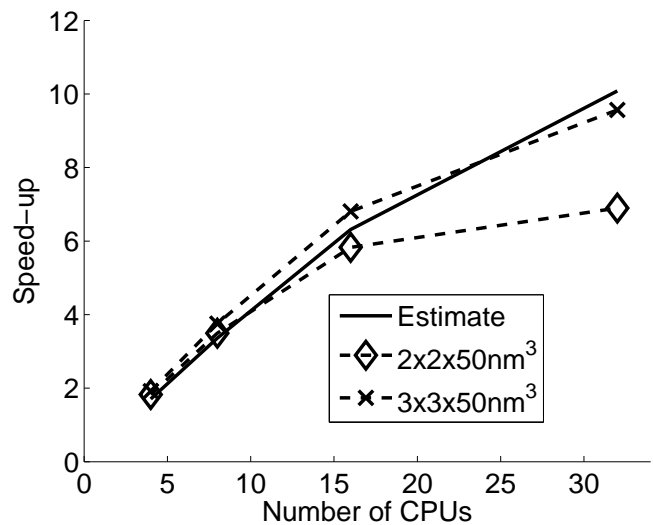}}                
  \quad\subfigure[
  $G^{<}$ speed-up is compared against theoretical estimate
from~(\ref{eqn:gless_speed}) for $100$nm length silicon nanowires.
  ]{\label{fig:gless_scale_100nm}\includegraphics[width=0.8\textwidth]{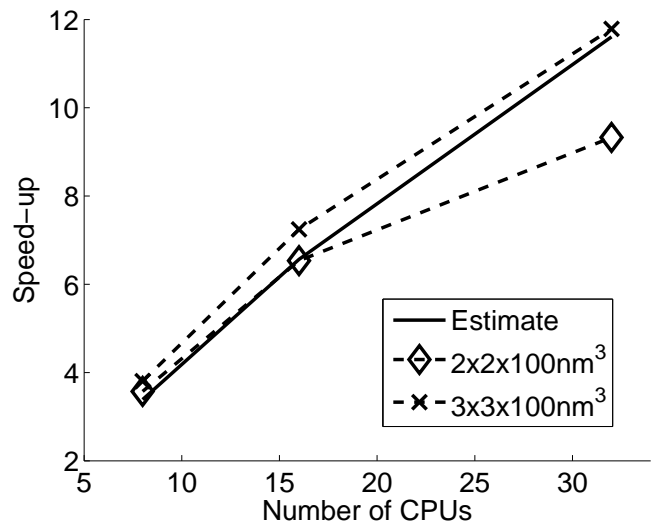}}
  \caption{  Verification of theoretical estimate for $G^{<}$ RGF speed-up considering 
  $50$nm and $100$nm length silicon nanowires.}
  \label{fig:gless_scale}
\end{figure}

Each NEGF nanowire simulation requires 
thousands of $G^{R}$, $G^{<}$, and $G^{>}$
 computations.
For our approach, as well as RGF,
the time for each computation 
will remain constant given that the underlying
structure of the Hamiltonian and scattering
matrices does not change.
In order to analyze the computational 
benefits of our approach we have 
examined several different cross-section
sizes and lengths of nanowires.
Specifically, 
silicon $\left[100\right]$ nanowires with lengths $35$nm, $53$nm,
and $70$nm are examined with cross-sections of $2$nm, $3$nm, and $4$nm.
Table~\ref{tab:gless_scale} shows the runtime for
the $G^{R}$ and $G^{<}$ computations (the time needed
for $G^{>}$ is identical to that of $G^{<}$).
''RGF$\times$'' is the observed speed-up for the combined time
of $G^{R}$, $G^{<}$, and $G^{>}$ calculations compared
to RGF.
''N/A'' is used to for devices that are not long 
enough to be divided based upon the number of
CPUs.
We can clearly see that
the efficiency of our algorithm improves
 as both the length and 
cross-section of the device increase.
There are two reasons behind this trend.  First,
considering a fixed number of processors a
longer device will devote less of the total 
time to sub-problem combining.
In addition, as the cross-section
size increases the inter-processor
communication costs
will be a smaller fraction of the total
simulation time.
We see these effects again when 
examining Figure~\ref{fig:gless_scale}.
Here, we verify the accuracy of the estimated
 $G^{<}$ scaling trend that
was derived in~(\ref{eqn:gless_speed}).
The speed-up over RGF for both 
$50$nm and $100$nm nanowires
are compared against our theoretical
estimate.
It is important
to note that although the speed-up 
estimate~(\ref{eqn:gless_speed})
is independent of the cross-section
size, effects such as data access time,
vector scaling/addition, and inter-processor 
communication will 
play a role in determining the efficiency.
Thus, in both Figures~\ref{fig:gless_scale_50nm}
and~\ref{fig:gless_scale_100nm} we again
see improved efficiency when considering the 
larger $3$nm cross-section nanowires.
In the case of the $3\times3\times100$nm$^3$
nanowire we achieve $11.8\times$ speed-up
when utilizing $32$ CPUs.

In addition to providing 
computational improvements,
our algorithm facilitates the simulation 
of larger cross-section devices.
If we consider the $4$nm cross-section
devices analyzed in 
Table~\ref{tab:gless_scale}, the
RGF method would require between
$16$GB and $32$GB of memory 
for lengths of $35$nm to $70$nm.
These devices would not be able to 
be analyzed without the use of special
purpose hardware.
\begin{figure}
\begin{center}
\leavevmode
\epsfxsize=3.2in
\epsfbox{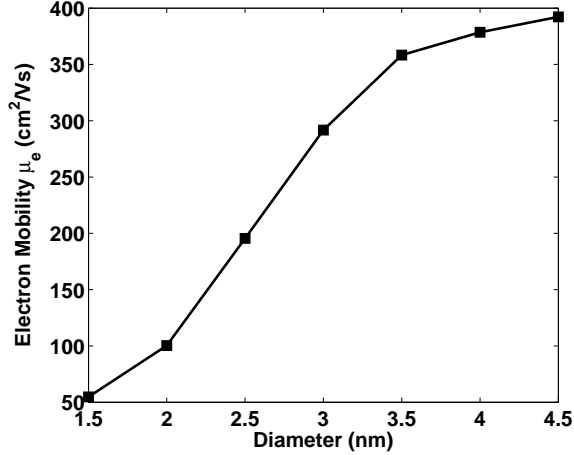}
\caption{Phonon-limited electron mobility $\mu_{ph}$ as function of the diameter
of $\left[100\right]$ oriented circular nanowires at room temperature. The electron
density along the channel is homogeneous and set to $n=1e20 cm^{-3}$.}
\label{fig:mobility}
\end{center}
\end{figure}
As an illustration of the capacity for our 
algorithm to simulate devices previously viewed 
to have prohibitive memory requirements, 
we stably generate the mobility 
for $4.5$nm cross-section devices
with electron-phonon scattering.  
In order to facilitate complete nanowire simulations 
we have implemented our algorithm on 
dual hex-core AMD Opteron 2435 (Istanbul) processors running at
 2.6GHz, 16GB of DDR2-800 memory, and a SeaStar 2+ router.
As an application, the low-field phonon-limited
 mobility of electrons
$\mu_{ph}$ is calculated in circular nanowires 
with diameters ranging
from $1.5$ to $4.5$nm and transport along the $\left[100\right]$
 crystal axis. The
"dR/dL" method~\cite{Rim_02} and the same procedure as
 in~\cite{mLU_11} are used to
obtain $\mu_{ph}$. The channel resistance "R" 
is computed as function of
the nanowire length "L" and then converted into 
a mobility. Here, "L"
is set to $35$nm, "R" is computed in the limit of
 ballistic transport
and in the presence of electron-phonon scattering,
 and the difference
between these two points is considered to evaluate
 dR/dL. The results
are shown in Figure~\ref{fig:mobility}. From a numerical perspective,
 the computation of
each Green's Function, at a given energy, was
 parallelized on 16 CPUs
for all the device structures.
As was alluded to above the simulation of 
these structures would not have been 
possible (due to memory restrictions)
without the decomposition
of the device through our 
parallel methods.

\section{Conclusions}
\label{ss:con}

In this work we have developed 
algorithms for parallel NEGF 
simulation with scattering.
The computational benefits of 
our approach have been 
demonstrated on large cross-section
silicon nanowires.  We show improvements
of over $11\times$ for the $G^{<}$
and $G^{>}$ computations.  In addition,
our approach enables simulations without
the need for special purpose hardware.
This can best be observed through our
simulation results for $4.5$nm cross-section
silicon nanowires.
The algorithms developed in this work
are applicable for a wide range of 
device geometries considering both 
atomistic and effective-mass models.
In addition to offering
significant computational improvements
 over the serial Recursive Green's Function
algorithm,
our approach facilitates simulation of
realistically sized devices on 
typical distributed computing hardware.


\appendix

\section{Parallel Computation of the Retarded Green's Function}
\label{ap:GR}

In Appendix~\ref{ap:GR} we build 
upon the approach of~\cite{cauley:07} 
to consider the computation of a distributed 
generator representation for $G^R$.
In Section~\ref{ss:Maps} we introduce a mapping
framework for combining sub-matrices corresponding
to subsets of the device structure.
This has similarities to the approach presented
in~\cite{cauley:07}, where in that case
only the diagonal entries of $G^R$ were needed
to describe the density of states.
Section~\ref{ss:recur_comb} illustrates how
a recursive process can be formulated in order
to reconstruct the compact representation of $G^R$.
This involves several key differences when 
compared to~\cite{cauley:07} as significantly 
more information is required from $G^R$.
Finally, in Section~\ref{ss:up_scheme} the 
parallel $G^R$ algorithm is summarized including
an analysis of the computational complexity.

\subsection{Matrix Maps}
\label{ss:Maps}

Matrix mappings are constructed in order to eventually 
produce the block tridiagonal portion of
$G^R$ while avoiding any unnecessary computation during the combining process.
Specifically, we will show that both the boundary 
block entries (first block row and last block column)
and the block tridiagonal entries from any combined inverse $\phi_{i\sim j}^{-1}$
must be attainable (not necessarily computed) for all combining steps.
We begin by illustrating the initial stage of the combining process given four
 divisions, where for simplicity each will be assumed to 
have $N$ blocks of size $N_x$.
First, the two sub-matrices $\phi_{1}$ and $\phi_{2}$ are connected through the
bridge matrix $B_{N}$ and together they form the larger block 
tridiagonal matrix $\phi_{1\sim2}$. 
By examining Figure~\ref{fig:decomp_over} it can be seen that eventually  
$\phi_{1\sim 2}^{-1}$ and $\phi_{3\sim 4}^{-1}$ will be combined and 
we must therefore produce the boundaries for each combined inverse.
From (\ref{eqn:minv}) the first block row and last block
 column of $\phi_{1\sim2}^{-1}$ can be calculated through the
use of an ``adjustment'' matrix:

\begin{align*}
J &= \begin{pmatrix} I&-\phi_{2}^{-1}\left(1,1\right)B_{N}^{T}\\
-\phi_{1}^{-1}\left(N,N\right)B_{N}&I \end{pmatrix}^{-1},
\end{align*}

\noindent as follows:

\begin{align}
\phi_{1\sim2}^{-1}(1,:) &=
\begin{pmatrix} \phi_{1}^{-1}(1,:) & 0 \end{pmatrix} - 
\begin{pmatrix} 
 \left[ -\phi_{1}^{-1}(1,N)B_{N}J_{12}\phi_{1}^{-1}(:,N)^{T} \right]^T\\
\left[ -\phi_{1}^{-1}(1,N)B_{N}J_{11}\phi_{2}^{-1}(1,:) \right]^T\\
 \end{pmatrix}^T,
\nonumber  \\
\label{eqn:rc_map}
\\ \nonumber
\phi_{1\sim2}^{-1}(:,2N) &=
\begin{pmatrix} 0 & \phi_{2}^{-1}(N,:) \end{pmatrix}^{T} - 
\begin{pmatrix}
\left[ -\phi_{2}^{-1}(N,1)B_{N}^{T}J_{22}\phi_{1}^{-1}(:,N)^{T} \right]^T \\
\left[ -\phi_{2}^{-1}(N,1)B_{N}^{T}J_{21}\phi_{2}^{-1}(1,:) \right]^T \\
 \end{pmatrix}. \nonumber 
\end{align}

\noindent In addition, the $r^{\mathrm{th}}$ diagonal block of $\phi_{1\sim2}^{-1}$ can be calculated 
using the following relationships for $r \leq N$ :

\begin{align}
\label{eqn:diag_map}
&\phi_{1\sim2}^{-1}(r,r) = \phi_{1}^{-1}(r,r) - 
\begin{pmatrix}
  -\phi_{1}^{-1}(r,N)B_{N}J_{12}\phi_{1}^{-1}(r,N)^{T}\end{pmatrix}, \nonumber \\ \\
&\phi_{1\sim2}^{-1}(r+N,r+N) = \phi_{2}^{-1}(r,r) - 
\begin{pmatrix}
  -\phi_{2}^{-1}(r,1)B_{N}^{T}J_{21}\phi_{2}^{-1}(1,r))\end{pmatrix}^{T}, \nonumber
\end{align}

\noindent  where the $r^{\mathrm{th}}$ off-diagonal block of $\phi_{1\sim2}^{-1}$ 
can be calculated using the following relationships:

\begin{align}
\label{eqn:odiag_map}
&\phi_{1\sim2}^{-1}(r,r+1) = \phi_{1}^{-1}(r,r+1) -  
\begin{pmatrix}
  -\phi_{1}^{-1}(r,N)B_{N}J_{12}\phi_{1}^{-1}(r+1,N)^{T}\end{pmatrix}, \nonumber \\
\\
&\phi_{1\sim2}^{-1}(r+N,r+1+N) = \phi_{2}^{-1}(r,r+1) -  
 \begin{pmatrix}
  -\phi_{2}^{-1}(r+1,1)B_{N}^{T}J_{21}\phi_{2}^{-1}(1,r))\end{pmatrix}^{T}, \nonumber \\
& r < N, \nonumber \\
 \nonumber
\\ \nonumber
&\phi_{1\sim2}^{-1}(r,r+1) = 0 - 
\begin{pmatrix}
  -\phi_{1}^{-1}(r,N)B_{N}J_{11}\phi_{2}^{-1}(1,1)\end{pmatrix}, \nonumber \\ 
& r = N. \nonumber 
\end{align}

\noindent The combination of $\phi_{3}$ and $\phi_{4}$ through the bridge matrix $B_{3N}$
results in similar relationships to those seen above.
Thus, in order be able to produce both the boundary and block 
tridiagonal portions of each
combined inverse we assign a total of twelve $N_{x} \times N_{x}$ matrix 
maps for each sub-matrix $k$.
$M_{k;1-4}$ describe effects for the $k^{\mathrm{th}}$ portion of the boundary,
$M_{k;5-8}$ describe the effects for a majority of the tridiagonal blocks, 
while $C_{k;1-4}$, which we will refer
to as ``cross'' maps, can be used to produce the remainder 
of the tridiagonal blocks.

Initially, for each sub-matrix $i$ the mappings $M_{k;i}  = I,~~k = 1,4,$ 
with all remaining mapping terms set to zero.  This ensures that initially the
 boundary of $\phi_{i\sim i}^{-1}$ matches the 
actual entries from the sub-matrix inverse,
 and the modifications to the tridiagonal portion due to combining
 are all set to zero.
By examining the first block row, last block column, and
the tridiagonal portion of the combined inverse $\phi_{1 \sim 2}^{-1}$ 
we can see how the maps can be used to explicitly represent all of the
needed information.
The governing responsibilities of the individual matrix maps are detailed
below:
\begin{align}
&\phi_{1 \sim 2}^{-1}(1,:) =  \begin{pmatrix} 
\left[ M_{1;1}\phi_{1}^{-1}(1,:) + M_{2;1}\phi_{1}^{-1}(:,N)^{T} \right]^T \\ \nonumber
\left[ M_{1;2}\phi_{2}^{-1}(1,:) + M_{2;2}\phi_{2}^{-1}(:,N)^{T} \right]^T \\ \nonumber
 \end{pmatrix}^T,
\nonumber \\
&\phi_{1 \sim 2}^{-1}(:,2N) = \begin{pmatrix} 
\left[ M_{3;1}\phi_{1}^{-1}(1,:) + M_{4;1}\phi_{1}^{-1}(:,N)^{T} \right]^T \\ \nonumber
\left[ M_{3;2}\phi_{2}^{-1}(1,:) + M_{4;2}\phi_{2}^{-1}(:,N)^{T} \right]^T\\ \nonumber
 \end{pmatrix}, 
\label{eqn:mmaps_wm}
\nonumber \\
&\phi_{1 \sim 2}^{-1}(r,s) = \phi_{1}^{-1}(r,s) - [ \phi_{1}^{-1}(r,1)M_{5;1}\phi_{1}^{-1}(1,s) +  
 \phi_{1}^{-1}(r,1)M_{6;1}\phi_{1}^{-1}(s,N)^{T} +\nonumber \\  
 & \phi_{1}^{-1}(r,N)M_{7;1}\phi_{1}^{-1}(1,s) + \phi_{1}^{-1}(r,N)M_{8;1}\phi_{1}^{-1}(s,N)^{T} ], \\  \nonumber
\nonumber \\
&\phi_{1 \sim 2}^{-1}(r,s+N) = - [ \phi_{1}^{-1}(r,1)C_{1;1}\phi_{2}^{-1}(1,s) + 
 \phi_{1}^{-1}(r,1)C_{2;1}\phi_{2}^{-1}(s,N)^{T} +\nonumber \\  
&\phi_{1}^{-1}(r,N)C_{3;1}\phi_{2}^{-1}(1,s) + \phi_{1}^{-1}(r,N)C_{4;1}\phi_{2}^{-1}(s,N)^{T} ], 
\nonumber 
\\ \nonumber
\nonumber \\
\nonumber
&\phi_{1 \sim 2}^{-1}(r+N,s+N) = \phi_{2}^{-1}(r,s) - [ \phi_{2}^{-1}(r,1)M_{5;2}\phi_{2}^{-1}(1,s) + 
 \phi_{2}^{-1}(r,1)M_{6;2}\phi_{2}^{-1}(s,N)^{T} +\nonumber \\  
&\phi_{2}^{-1}(r,N)M_{7;2}\phi_{2}^{-1}(1,s) + \phi_{2}^{-1}(r,N)M_{8;2}\phi_{2}^{-1}(s,N)^{T} ], 
\nonumber \\ \nonumber
\\ \nonumber
& r,s \leq N, 
\nonumber
\end{align}

\begin{figure}
\begin{center}
\leavevmode
\epsfxsize=3.6in
\epsfbox{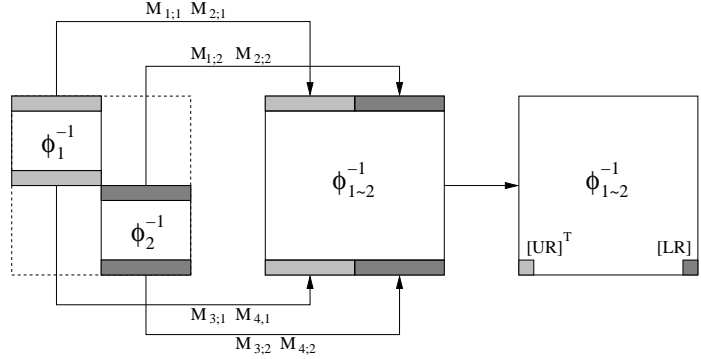}
\caption{Mapping dependencies when combining $\phi_{1}^{-1}$ and 
$\phi_{2}^{-1}$ to form $\phi_{1 \sim 2}^{-1}$.}
\label{fig:comb_maps}
\end{center}
\end{figure}
\noindent It is important to note that all of the 
expressions~(\ref{eqn:rc_map})-(\ref{eqn:odiag_map})
can be written into the matrix map framework of~(\ref{eqn:mmaps_wm}).
Figure~\ref{fig:comb_maps} shows the mapping dependencies for the first 
block row and last block row
 (or column since $K$ is symmetric).  From~(\ref{eqn:rc_map}) we see that 
both of the block rows are distributed based 
upon the location of each sub-matrix with respect to the 
bridge point, i.e. the mapping terms associated with $\phi_{1}^{-1}$ can be used to 
produce the first portion of the rows while those associated with $\phi_{2}^{-1}$
can be used for the remainder.
In fact, this implicit division for the mapping dependencies holds for the block
tridiagonal portion of the combined inverses as well, enabling an efficient parallel
 implementation.
Thus, from this point we can deduce that the matrix maps for the 
first block row~(\ref{eqn:mmaps_wm}) must be updated in the following manner:

\begin{align*}
M_{1;1} \leftarrow M_{1;1}+(\phi_{1}^{-1}(1,N) B_{N} J_{12} )M_{3;1};
\nonumber \\ 
M_{2;1} \leftarrow M_{2;1}+(\phi_{1}^{-1}(1,N) B_{N} J_{12} )M_{4;1};
\\
M_{3;1} \leftarrow (\phi_{1}^{-1}(1,N) B_{N} J_{11})M_{1;2};
\nonumber \\
M_{4;1} \leftarrow (\phi_{1}^{-1}(1,N) B_{N} J_{11})M_{2;2};
\end{align*}

In order to understand these relationships 
it is important to first recall that the updates to the maps
associated with sub-matrix $\phi_{1}$ are dependent on the last block 
column $\phi_{1}^{-1}(:,N)$.  Thus, we see a dependence
on the previous state for the last block column $\phi_{1}^{-1}(:,N)$, 
i.e. the new state of the mapping terms $M_{1;1}$ and $M_{2;1}$  
are dependent on the previous state of the mapping terms $M_{3;1}$ and $M_{4;1}$
respectively.
Similarly, a dependence on $\phi_{2}^{-1}(1,:)$ results in  
the new state of the mapping terms $M_{1;2}$ and $M_{2;2}$  
being functions of the previous state of the mapping terms $M_{1;2}$ and $M_{1;2}$
respectively.
Finally, although some of the mapping terms remain zero after 
this initial combining step ($M_{2;2}$ for example), 
the expressions described in~(\ref{eqn:mmaps_wm})
need to be general enough for the methodology.   That is, the mapping 
expressions must be able to capture combining effects for multiple
combing stages, regardless of the position of the sub-matrix with
 respect to a bridge point.
For example, if we consider sub-matrix $\phi_2$ for the case
 seen in Figure~\ref{fig:decomp_over},
during the initial combining step it would be considered a lower problem
and for the final combining step it would be considered a upper problem.  Alternatively,
sub-matrix $\phi_{3}$ would be associated with exactly the opposite modifications.
It is important to note that every possible modification process, for 
the individual mapping terms, is encompassed within this
 general matrix map framework.

\subsection{Recursive Combining Process}
\label{ss:recur_comb}

In order to formalize the notion of a recursive update scheme
we will continue the example from Section~\ref{ss:Maps}.
By examining the final combing stage for the case of four divisions,
we notice that the approach described in~(\ref{eqn:rc_map})-(\ref{eqn:odiag_map})
can again be used to combine sub-matrix inverses $\phi_{1 \sim 2}^{-1}$ and $\phi_{3 \sim 4}^{-1}$,
through the bridge matrix $B_{2N}$.
The first block row and last block
 column of $\phi_{1\sim4}^{-1}$ can be calculated as follows:
\begin{align}
\phi_{1\sim4}^{-1}(1,:) &=
\begin{pmatrix} \phi_{1\sim2}^{-1}(1,:) & 0 \end{pmatrix} - 
\begin{pmatrix} 
 \left[ -\phi_{1\sim2}^{-1}(1,2N)B_{2N}J_{12}\phi_{1\sim2}^{-1}(:,2N)^{T} \right]^T\\
\left[ -\phi_{1\sim2}^{-1}(1,2N)B_{2N}J_{11}\phi_{3\sim4}^{-1}(1,:) \right]^T\\
 \end{pmatrix}^T,
\nonumber  \\
\label{eqn:rc_map_2lvl}
\\ \nonumber
\phi_{1\sim4}^{-1}(:,4N) &=
\begin{pmatrix} 0 & \phi_{3\sim4}^{-1}(2N,:) \end{pmatrix}^{T}  - 
\begin{pmatrix}
\left[ -\phi_{3\sim4}^{-1}(2N,1)B_{2N}^{T}J_{22}\phi_{1\sim2}^{-1}(:,2N)^{T} \right]^T \\
\left[ -\phi_{3\sim4}^{-1}(2N,1)B_{2N}^{T}J_{21}\phi_{3\sim4}^{-1}(1,:) \right]^T \\
 \end{pmatrix}, \nonumber 
\end{align}
given the adjustment matrix:

\begin{align*}
J &= \begin{pmatrix} I&-\phi_{3\sim4}^{-1}\left(1,1\right)B_{2N}^{T}\\
-\phi_{1\sim2}^{-1}\left(2N,2N\right)B_{2N}&I \end{pmatrix}^{-1}.
\end{align*}

In addition, the $r^{\mathrm{th}}$ diagonal block of $\phi_{1\sim4}^{-1}$ can be calculated 
using the following relationships:
\begin{align}
&\phi_{1\sim4}^{-1}(r,r) = \phi_{1\sim2}^{-1}(r,r) - 
\begin{pmatrix}
  -\phi_{1\sim2}^{-1}(r,2N)B_{2N}J_{12}\phi_{1\sim2}^{-1}(r,2N)^{T}\end{pmatrix}, \nonumber \\
\label{eqn:diag_map_2lvl}
\\ \nonumber
&\phi_{1\sim4}^{-1}(r+2N,r+2N) = \phi_{3\sim4}^{-1}(r,r) -  
\begin{pmatrix}
  -\phi_{3\sim4}^{-1}(r,1)B_{2N}^{T}J_{21}\phi_{3\sim4}^{-1}(1,r))\end{pmatrix}^{T},  \nonumber \\
& r \leq 2N, \nonumber
\end{align}

\noindent where the $r^{\mathrm{th}}$ off-diagonal block of $\phi_{1\sim4}^{-1}$ 
can be calculated using the following relationships:
\begin{align}
\label{eqn:odiag_map_2lvl}
&\phi_{1\sim4}^{-1}(r,r+1) = \phi_{1\sim2}^{-1}(r,r+1) -  
 \begin{pmatrix}
  -\phi_{1\sim2}^{-1}(r,2N)B_{2N}J_{12}\phi_{1\sim2}^{-1}(r+1,2N)^{T}\end{pmatrix}, \nonumber \\
\\
&\phi_{1\sim4}^{-1}(r+2N,r+1+2N) = \phi_{3\sim4}^{-1}(r,r+1) -  
 \begin{pmatrix}
  -\phi_{3\sim4}^{-1}(r+1,1)B_{2N}^{T}J_{21}\phi_{3\sim4}^{-1}(1,r))\end{pmatrix}^{T}, \nonumber \\
& r < 2N, \nonumber \\
 \nonumber
\\ \nonumber
&\phi_{1\sim4}^{-1}(r,r+1) = 0 - 
\begin{pmatrix}
  -\phi_{1\sim2}^{-1}(r,2N)B_{2N}J_{11}\phi_{3\sim4}^{-1}(1,1)\end{pmatrix}, \nonumber \\ 
& r = 2N. \nonumber 
\end{align}

\noindent Again, it is important to note that each of the 
expressions~(\ref{eqn:rc_map_2lvl})-(\ref{eqn:odiag_map_2lvl})
are implicitly divided based upon topology.
For example, the first $2N$ diagonal blocks
 of $\phi_{1\sim4}^{-1} = G^R$ can be separated into
two groups based upon the size of the sub-matrices $\phi_1$ and 
$\phi_2$.  That is,
\begin{align*}
&\phi_{1\sim4}^{-1}(r,r) = \phi_{1\sim2}^{-1}(r,r) - 
\begin{pmatrix}
  -\phi_{1\sim2}^{-1}(r,2N)B_{2N}J_{12}\phi_{1\sim2}^{-1}(r,2N)^{T}\end{pmatrix}, \nonumber \\
& r \leq 2N,
\end{align*}
\noindent can be separated for $ r \leq N$ as:
\begin{align*}
&\phi_{1\sim4}^{-1}(r,r) = \phi_{1}^{-1}(r,r) - 
\begin{pmatrix}
\left[ \phi_{2}^{-1}(N,1)B_{N}^{T}J_{22}\phi_{1}^{-1}(r,N)^{T} \right]^T \\
 \end{pmatrix} \cdot B_{2N}J_{12} \cdot \\
&\begin{pmatrix}
\left[ \phi_{2}^{-1}(N,1)B_{N}^{T}J_{22}\phi_{1}^{-1}(r,N)^{T} \right] \\
\end{pmatrix}, \\
\\
&\phi_{1\sim4}^{-1}(r+N,r+N) = \phi_{2}^{-1}(r,r) - 
\begin{pmatrix}
\left[ \phi_{2}^{-1}(N,r)+\phi_{2}^{-1}(N,1)B_{N}^{T}J_{21}\phi_{2}^{-1}(1,r) \right]^T \\
 \end{pmatrix} \cdot B_{2N}J_{12} \cdot \\
&\begin{pmatrix}
\left[ \phi_{2}^{-1}(N,r)+\phi_{2}^{-1}(N,1)B_{N}^{T}J_{21}\phi_{2}^{-1}(1,r) \right] \\
\end{pmatrix}.
\end{align*}
\noindent Thus, the modifications to the diagonal entries can be written 
as just a function of the first block row and last block column from the individual
 sub-matrices, using the matrix map framework introduced in~(\ref{eqn:mmaps_wm}) for
$ r \leq N$ :
\begin{align*}
&\phi_{1\sim4}^{-1}(r,r) = \phi_{1}^{-1}(r,r) - 
\begin{pmatrix}
\left[ M_{1;1}\phi_{1}^{-1}(1,r) + M_{2;1}\phi_{1}^{-1}(r,N)^{T} \right]^T \\
 \end{pmatrix} \cdot B_{2N}J_{12} \cdot \\
&\begin{pmatrix}
\left[ M_{1;1}\phi_{1}^{-1}(1,r) + M_{2;1}\phi_{1}^{-1}(r,N)^{T} \right] \\
\end{pmatrix}, \\
\\
&\phi_{1\sim4}^{-1}(r+N,r+N) = \phi_{2}^{-1}(r,r) - 
\begin{pmatrix}
\left[ M_{1;2}\phi_{2}^{-1}(1,r) + M_{2;2}\phi_{2}^{-1}(r,N)\right]^T \\
 \end{pmatrix} \cdot B_{2N}J_{12} \cdot \\
&\begin{pmatrix}
\left[ M_{1;2}\phi_{2}^{-1}(1,r) + M_{2;2}\phi_{2}^{-1}(r,N) \right] \\
\end{pmatrix}. 
\end{align*}
\noindent Here, the matrix maps are assumed to have been updated based upon the
formation of the combined inverses $\phi_{1\sim2}^{-1}$ and $\phi_{3\sim4}^{-1}$.
Therefore, we can begin to formulate the recursive framework for updating
the matrix maps to represent the effect of each combining step.  

\subsection{Update Scheme for Parallel Inversion and the Distributed Compact Representation}
\label{ss:up_scheme}

The procedure begins with each division of the problem being assigned to one of $p$ available
CPUs.  In addition, all of the $p-1$ bridge matrices are made available to each of the CPUs.
After the compact representation for each inverse has been found independently, the combining
process begins.
  Three reference positions are defined for the formation of 
a combined inverse $\phi_{i \sim j}^{-1}$: the ``start'' position $[\mathtt{st}] = i$, 
the ``stop'' position $[\mathtt{sp}] = j$,
 and the bridge position $[\mathtt{bp}] = \lceil \frac{j - i}{2}  \rceil$. 
Due to the fact that a CPU $t$ will only be involved in the formulation 
of a combined inverse when $[\mathtt{st}] \leq t \leq [\mathtt{sp}]$ all 
combining stages on the same level (see Figure~\ref{fig:decomp_over}) 
can be performed concurrently.
When forming a combined inverse $\phi_{i \sim j}^{-1}$,
each CPU $[\mathtt{st}] \leq t \leq [\mathtt{sp}]$ will first need to
form the adjustment matrix for the combining step.
Assuming a bridge matrix $B_{k}$, we begin by constructing 
four ``corner blocks''.
If the upper combined inverse is assumed to have
$N_u$ blocks and the lower to have $N_{l}$,
the two matrices need from the upper combined inverse are:
$[\mathtt{UR}] = \phi_{[\mathtt{st}]\sim[\mathtt{bp}]}^{-1}(1,N_u)
~\mathrm{and}~[\mathtt{LR}] = \phi_{[\mathtt{st}]\sim[\mathtt{bp}]}^{-1}(N_u, N_u)$,
with the two matrices from the lower being:
$[\mathtt{UL}] = \phi_{[\mathtt{bp+1}]\sim[\mathtt{sp}]}^{-1}(1,1)
~\mathrm{and}~[\mathtt{LL}] = \phi_{[\mathtt{bp+1}]\sim[\mathtt{sp}]}^{-1}(N_l,1)$.
These matrices can be generated by the appropriate CPU
through their respective matrix maps (recall the example shown in Figure~\ref{fig:comb_maps}).
  Specifically, the CPUs corresponding to the 
$[\mathtt{st}],~[\mathtt{bp}],~[\mathtt{bp+1}]~\mathrm{and}~[\mathtt{sp}]$  
divisions govern the required information.
The adjustment matrix for the combining step can then be formed:

\begin{align*}
J &= \begin{pmatrix} I&-[\mathtt{UL}]B_{k}^{T}\\
-[\mathtt{LR}]B_{k}&I \end{pmatrix}^{-1}.
\end{align*}

\noindent After the adjustment matrix has been calculated the process of updating the matrix
maps can begin.
For any combining step, the cross maps each CPU $t$ must be updated first:

\begin{align}
&\mathrm{if}~~(t < [\mathtt{bp}])~~\mathrm{then} \nonumber \\
&~~C_{1}\leftarrow C_{1}-M_{3;t}^{T}(B_{k} J_{12})M_{3;t+1};
\nonumber \\
&~~C_{2}\leftarrow C_{2}-M_{3;t}^{T}(B_{k} J_{12})M_{4;t+1};
\nonumber \\
&~~C_{3}\leftarrow C_{3}-M_{4;t}^{T}(B_{k} J_{12})M_{3;t+1};
\nonumber \\
&~~C_{4}\leftarrow C_{4}-M_{4;t}^{T}(B_{k} J_{12})M_{4;t+1};
\nonumber \\
&\mathrm{else if}~~(t == [\mathtt{bp}])~~\mathrm{then} 
\label{eqn:maps_c}  \\
&~~C_{1}\leftarrow C_{1}-M_{3;t}^{T}(B_{k} J_{11})M_{1;t+1};
\nonumber \\
&~~C_{2}\leftarrow C_{2}-M_{3;t}^{T}(B_{k} J_{11})M_{2;t+1};
\nonumber \\
&~~C_{3}\leftarrow  C_{3}-M_{4;t}^{T}(B_{k} J_{11})M_{1;t+1};
\nonumber \\
&~~C_{4}\leftarrow C_{4}-M_{4;t}^{T}(B_{k} J_{11})M_{2;t+1};
\nonumber \\
&\mathrm{else if}~~(t < [\mathtt{sp}])~~\mathrm{then} \nonumber \\
&~~C_{1}\leftarrow C_{1}-M_{1;t}^{T}(B_{k} ^ {T} J_{21})M_{1;t+1};
\nonumber \\
&~~C_{2}\leftarrow C_{2}-M_{1;t}^{T}(B_{k} ^ {T} J_{21})M_{2;t+1};
\nonumber \\
&~~C_{3}\leftarrow C_{3}-M_{2;t}^{T}(B_{k} ^ {T} J_{21})M_{1;t+1};
\nonumber \\
&~~C_{4}\leftarrow C_{4}-M_{2;t}^{T}(B_{k} ^ {T} J_{21})M_{2;t+1};
\nonumber
\end{align}

\noindent Notice that the cross maps for CPU $t$ are dependent on 
information from its neighboring CPU $t+1$.  This information must
be transmitted and made available before the cross updates can be 
performed.
Next, updates to the remaining eight matrix maps can be separated into
two categories.
The updates to the matrix maps for the upper sub-matrices $(t \leq [\mathtt{bp}])$, 
are summarized below:
\begin{align}
&M_{5;t}\leftarrow M_{5;t}-M_{3;t}^{T}(B_{k}J_{12} )M_{3;t};
\nonumber \\   
&M_{6;t}\leftarrow M_{6;t}-M_{3;t}^{T}(B_{k}J_{12} )M_{4;t};
\nonumber \\
&M_{7;t}\leftarrow M_{7;t}-M_{4;t}^{T}(B_{k}J_{12} )M_{3;t};
\nonumber \\   
&M_{8;t}\leftarrow M_{8;t}-M_{4;t}^{T}(B_{k}J_{12} )M_{4;t};
\label{eqn:maps_u}  \\
&M_{1;t}\leftarrow M_{1;t}+([\mathtt{UR}] B_{k} J_{12} )M_{3;t};
\nonumber \\   
&M_{2;t}\leftarrow M_{2;t}+([\mathtt{UR}] B_{k} J_{12} )M_{4;t};
\nonumber \\   
&M_{3;t}\leftarrow ([\mathtt{LL}]^{T} B_{k}^{T} J_{22})M_{3;t};
\nonumber \\   
&M_{4;t}\leftarrow ([\mathtt{LL}]^{T} B_{k}^{T} J_{22})M_{4;t};
\nonumber 
\end{align}

\noindent The updates to the matrix maps for the lower 
sub-matrices $(t > [\mathtt{bp}])$, will be:
\begin{align}
&M_{5;t}\leftarrow M_{5;t}-M_{1;t}^{T}(B_{k}^{T}J_{21})M_{1;t};
\nonumber \\
&M_{6;t}\leftarrow M_{6;t}-M_{1;t}^{T}(B_{k}^{T}J_{21})M_{2;t};
\nonumber \\
&M_{7;t}\leftarrow M_{7;t}-M_{2;t}^{T}(B_{k}^{T}J_{21})M_{1;t};
\nonumber \\
&M_{8;t}\leftarrow M_{8;t}-M_{2;t}^{T}(B_{k}^{T}J_{21})M_{2;t};
\label{eqn:maps_l} \\ 
&M_{3;t}\leftarrow M_{3;t}+([\mathtt{LL}]^{T} B_{k}^{T} J_{21})M_{1;t};
\nonumber \\
&M_{4;t}\leftarrow M_{4;t}+([\mathtt{LL}]^{T} B_{k}^{T} J_{21})M_{2;t};
\nonumber \\
&M_{1;t}\leftarrow ([\mathtt{UR}] B_{k} J_{11})M_{1;t};
\nonumber \\
&M_{2;t}\leftarrow ([\mathtt{UR}] B_{k} J_{11})M_{2;t};
\nonumber
\end{align}

\noindent The above procedure, shown in~(\ref{eqn:maps_c})-(\ref{eqn:maps_l}), for modifying
 the matrix maps can be recursively repeated for each of the combining 
stages beginning with the lowest level of combining the individual sub-matrix inverses.

  On completion the maps can then be used to generate the block tridiagonal
entries of $G^R$.
This subsequently
allows for the computation of the generator sequences for $G^R$, via
the relationships shown in~(\ref{eqn:comp_ratio}), in a purely  
distributed fashion.
 The time complexity of the algorithm presented
is $O({N_{x}^{3}N_{y}}/{p} + N_{x}^{3}\log p)$, with memory consumption
$O\left({N_{x}^{2}N_{y}}/{p} + N_{x}^{2}\right)$.  The first term $({N_{x}^{3}N_{y}}/{p})$
in the computational complexity arises from the embarrassingly parallel nature of both determining
the generator sequences and applying the matrix maps to update the block tridiagonal portion
of the inverse.  The second term $(N_{x}^{3}\log p)$ is 
dependent on the number of levels needed to gather combining
information for $p$ sub-matrix inverses.  Similarly,
the first term in the memory complexity is due to the generator sequences and diagonal blocks, and
the second represents the memory required for the matrix maps of each sub-problem governed.

\end{document}